\documentstyle[preprint,12pt,aps,amssymb]{revtex}
\begin{document}

\title{Separation of variables and exact solution of the Klein-Gordon and Dirac equations
in an open universe}
\author{V\'{\i}ctor M. Villalba\thanks{On leave from Centro de F\'{\i}sica, 
Instituto Venezolano de Investigaciones Cient\'{\i}ficas, IVIC 
Apdo 21827, Caracas 1020-A, Venezuela.
e-mail: villalba@th.physik.uni-frankfurt.de}\footnote{Alexander von Humboldt Fellow}} 
\address{Institut f\"ur Theoretische Physik. Universit\"at Frankfurt \\
D-60054 Frankfurt am Main, Germany}
\author{Esteban Isasi Catal\'a\footnote{eisasi@fis.usb.ve}}
\address{Departamento de Fisica, Universidad Sim\'on Bol\'{\i}var, \\
Caracas, 89000 Venezuela.}

\maketitle

\begin{abstract}
We solve the Klein-Gordon and Dirac equations in an open cosmological universe
with a partially horn topology in the presence of a time dependent magnetic
field. Since the exact solution cannot be obtained explicitly for arbitrary
time-dependence of the field, we discuss the asymptotic behavior of the 
solutions with the help of the relativistic Hamilton-Jacobi equation. 
\end{abstract}


\section{Introduction}

During the last years a large amount of observational data has been reported
showing that our universe is almost isotropic and homogeneous. 
The study of the structure of the Cosmic
Microwave Radiation leads us to conclude that the ratio of the total density
to the critical density of the universe $\Omega _{0}$ is likely to be close
to one \cite{Bernardis,Perlmutter1,Perlmutter2}, favoring a spatially flat
Robertson-Walker metric over other topologies.

It is well known that general relativity is a local metrical theory and
therefore the corresponding Einstein field equations do not fix the global
topology of spacetime and consequently the universe may have compact spatial
sections with a nontrivial topology \cite{Barrow,Sokolov}, then the
observational data does not rule out the possibility that our universe
possesses a hyperbolic topology \cite{Barrow,Reboucas1,Reboucas2,Reboucas3}.

The study of cosmological models with nonstandard topologies is not new and
goes back to the works by Zelmanov \cite{Zelmanov,Vladimirov}, showing that
upon different coordinate transformations, spatially closed or flat sections
can be transformed into hyperbolic sections and vice versa.

The line element associated with an spatially open Friedman universe has the
form

\begin{equation}
ds^{2}=a^{2}(\eta )\left[ -d\eta ^{2}+dr^{2}+\sinh ^{2}\left( d\theta
^{2}+\sin ^{2}\theta d\phi ^{2}\right) \right]  \label{1}
\end{equation}
Making the coordinate transformation \cite{Sokolov} 
\begin{equation}
e^{-z}=\cosh r-\sinh r\cos \theta ,e^{-z}x=\sin \theta \cos \phi \sinh
r,e^{-z}y=\sin \theta \sin \phi \sinh r  \label{trans}
\end{equation}
the metric (\ref{1}) becomes

\begin{equation}
a^{-2}(\eta )ds^{2}=-d\eta ^{2}+dz^{2}+e^{-2z}\left( dx^{2}+dy^{2}\right)
\label{ani}
\end{equation}
The topology is induced by identifying points periodically along $x$ and $y$
by $(x,y)=(x+b,y+h)$, where $b$ and $h$ are constant, to create a two
dimensional torus. The torus is stretched by the factor $e^{-2z}$ along the $%
z$ axis to create a toroidal horn. The comoving proper area of the torus is $%
e^{-2z}bh$ and depends on location along the $z$ axis. The global topology
induces global inhomogeneity as well as global anisotropy \cite{Barrow}.

The study of quantum effects in cosmological universes with a nontrivial
topology allows us a deeper understanding of the properties of different
scenarios and which of them can describe our universe. In this direction the
metric (\ref{ani}) represents a very interesting scenario in order to
discuss particle production and propagation of perturbations in cosmology.

After the publication of the pioneering article by Schr\"{o}dinger \cite
{Schrodinger}, discussing particle production in a deSitter universe, many
articles have been published on the problem of quantum effects in
cosmological scenarios \cite{Birrel,Grib,Parker}, most of them dealing with a
Robertson-Walker line element with spatially flat topology. This
particularly simple line element, which is the most used in inflationary
models, permits one to compute the Green function as well as to solve the
relativistic wave equations \cite{Barut,Gitman,Parker71a}.

In order to study quantum processes in curved space-times one has to fulfill
a preliminary step which consists in having a description of the single-mode
solution of the relativistic particles or perturbations in those background
fields, i.e., exact solution of the relativistic scalar and spinor wave
equations. In the literature we have at our disposal different methods for
solving relativistic wave equations in curved spaces and in curvilinear
coordinates; among them the method of separation of variables is one of the
most widely used. \cite{Bagrov1,Bagrov2,Shishkin,Villalba1}.

It is the purpose of the present article to solve the Klein-Gordon and Dirac
equations in the Friedman universe associated with the metric (\ref{ani}) in
the presence of a time dependent magnetic field. In order to solve the Dirac
equation we apply the algebraic method of separation of variables \cite
{Shishkin,Villalba1,Villalba2,Hounkonnou,Hounkonnou2}. We compare the
solutions with those of obtained after solving the relativistic Hamilton
Jacobi equation. The article is structured as follows: In Sec. II we solve
the relativistic Hamilton-Jacobi equation in an open cosmological universe
with a horn topology. In Sec. III we separate variables and solve the
Klein-Gordon equation. In Sec IV, using the algebraic method of separation
of variables, we reduce the Dirac equation to a system of first order
coupled differential equations that we solve in terms of special functions.
Finally, in Sec. V we briefly discuss the results reported in this article.

\section{Solution of the Hamilton-Jacobi equation}

The covariant generalization of the Hamilton-Jacobi equation has the form \cite{Landau}
\begin{equation}
g^{\alpha \beta }\left( \frac{\partial S}{\partial x^{\alpha }}-eA_{\alpha
}\right) \left( \frac{\partial S}{\partial x^{\beta }}-eA_{\beta }\right)
+M^{2}=0,  \label{3}
\end{equation}
where $g^{\alpha \beta }$ is the contravariant metric, $A_{\alpha }$ is the
vector potential and $M$ is the mass of the particle. Here and elsewhere we
adopt the convention $c=\hbar =1$.

Let us introduce an electromagnetic field associated with the vector
potential

\begin{equation}  \label{A}
A^{\mu }=A_{1}(y)\delta _{1}^{\mu },
\end{equation}
where the index $\mu=0$ is associated with the evolution parameter $\eta$
and $\mu=1,2,3$ correspond to the space coordinates $x,y,z$ respectively. 
Looking at the relativistic invariants
\begin{equation}  \label{inv2}
\frac{1}{2}F^{\mu \nu }F_{\mu \nu }^{{}}=B^{2}-E^{2}=\frac{e^{4z}}{\alpha
(\eta )^{4}}\left( \frac{dA_{1}(y)}{dy}\right) ^{2},
\end{equation}
\begin{equation}  \label{inv1}
F^{\mu \nu }F_{\mu \nu }^{\ast }=0,
\end{equation}
and taking into account that only $F_{23}$ is different from zero, we
notice that (\ref{A}) corresponds to a non constant
magnetic field $B$, directed along the $z$ axis, with strength.
\begin{equation}
B=\frac{e^{2z}}{\alpha (\eta )^{2}}\left| \frac{dA_{1}(y)}{dy}\right|,  \label{B}
\end{equation}
whose value is inversely proportional to the expansion factor
$\alpha(\eta)^2$.

The line element (\ref{ani}) is a St\"{a}ckel space \cite{Stackel}, and the
Hamilton-Jacobi equation (\ref{3}) is completely separable in (\ref{ani}) in
the presence of the vector potential (\ref{A}), therefore we can look for a
solution in the form

\begin{equation}
S=k_{x}+S_{y}(y)+S_{z}(z)+S_{\eta }(\eta ).  \label{SHJ}
\end{equation}
Substituting (\ref{SHJ}) into Eq. (\ref{3}) we obtain 
\begin{equation}
\frac{(k_{x}-A_{1}(y))^{2}}{e^{-2z}}+\frac{1}{e^{-2z}}\left( \frac{dS_{y}}{dy%
}\right) ^{2}+\left( \frac{dS_{z}}{dz}\right) ^{2}-\left( \frac{dS_{\eta }}{%
d\eta }\right) ^{2}-M^{2}\alpha (\eta )^{2}=0.  \label{4}
\end{equation}
Equation (\ref{4}) reduces to the following system of differential equations: 
\begin{equation}
\left( \frac{dS_{z}}{dz}\right) ^{2}+k_{xy}^{2}e^{2z}=k_{z}^{2},  \label{5}
\end{equation}
\begin{equation}
\left( \frac{dS_{\eta }}{d\eta }\right) ^{2}+M^{2}\alpha (\eta
)^{2}=k_{z}^{2},  \label{6}
\end{equation}
\begin{equation}
(k_{x}-A_{1}(y))^{2}+\left( \frac{dS_{y}}{dy}\right) ^{2}=k_{xy}^{2},
\label{7}
\end{equation}
where $k_{xy}^{2},$ and $k_{z}^{2}$ are separation constants. 

In the absence of electromagnetic interaction, we have that $A_{1}(y)=0$ and
the solution of Eq. (\ref{7}) takes the form
\begin{equation}
\label{ky}
S_{y}=\pm \sqrt{k_{xy}^{2}-k_{x}^{2}}y=\pm k_{y}y.
\end{equation}
where we have introduced the constant $k_{y}$. Equation (\ref{ky}) can also be
derived looking at the symmetry between the torus coordinates $x$
and $y$ in the line element (\ref{ani}) and Eq. (\ref{4}) when $A_{1}(y)=0$.

When the vector potential has the simple form $A_{1}(y)=A_{1}y$, the magnetic
field reads $B=\frac{e^{2z}}{\alpha (\eta )^{2}}\left| A_{1}\right| $ and
the function \ $S_{y}(y)$ is

\begin{equation}  \label{9}
S_{y}(y)=-\frac{k_{x}-A_{1}y}{2A_{1}}\sqrt{k_{xy}^{2}-(k_{x}-A_{1}y)^{2}}+%
\frac{k_{xy}^{2}}{2A_{1}}\arctan \frac{A_{y}y-k_{x}}{\sqrt{%
k_{xy}^{2}-(k_{x}-A_{1}y)^{2}}},
\end{equation}
The solution of Eq. (\ref{5}) can be expressed in terms of elementary
functions as follows:

\begin{equation}
S_{z}=\sqrt{k_{z}^{2}-k_{xy}^{2}\exp (2z)}-k_{z}\tanh ^{-1}\sqrt{\frac{%
k_{z}^{2}-k_{xy}^{2}\exp (2z)}{k_{z}^{2}}}.  \label{8}
\end{equation}
The solution of Eq. (\ref{6}) can be written as 
\begin{equation}
S_{\eta }=\pm \int \sqrt{k_{z}^{2}-M^{2}\alpha (\eta )^{2}}d\eta,
\end{equation}
whose explicit form in terms of elementary functions will depend on a
particular choice of the expansion function $\alpha (\eta ) $.

Since we have been able to solve the Hamilton-Jacobi equation in the
St\"{a}ckel space given by (\ref{3}), we can construct the quasiclassical
modes of the relativistic wave equations through the identification 
\begin{equation}
\Phi \rightarrow e^{iS}=e^{\pm i\int \sqrt{k_{z}^{2}-M^{2}\alpha (\eta )^{2}}%
d\eta }e^{ik_{x}x}e^{iS_{y}}e^{iS_{z}},
\end{equation}
where $S_{z}$ and $S_{y}$ take the following values at the asymptotes 
\begin{equation}
S_{z(\infty )}\rightarrow ik_{xy}e^{z},\,S_{z(-\infty )}\rightarrow k_{z}z,
\label{Sz}
\end{equation}
\begin{equation}
S_{y(\infty )}\rightarrow \mp i\frac{(k_{x}-A_{1}y)^{2}}{2A_{1}}.  \label{Sy}
\end{equation}
When the electromagnetic interaction is not present we have that $S_{y}=\exp
(ik_{y}y)$.

\section{Solution of the Klein Gordon equation}

The covariant generalization of the Klein Gordon equation in curved
space-time has the form \cite{Birrel} 
\begin{equation}
g^{\alpha \beta }\left( \nabla _{\alpha }-ieA_{\alpha }\right) \left( \nabla
_{\beta }-ieA_{\beta }\right) \Phi -(M^{2}+\xi R)\Phi =0,  \label{KG}
\end{equation}
where $\newline
\nabla _{\alpha }$ is the covariant derivative, $R$ is the curvature scalar
and $\xi $ is a scalar dimensionless \ coupling constant which takes the
value $\xi =1/6$ in the conformal case and $\xi =0$ when a minimal coupling
is considered. The value of the $R$ for the metric (\ref{ani}) is 
\begin{equation}
R=6\frac{-a(\eta )+\frac{d^{2}a(\eta )}{d\eta ^{2}}}{a(\eta )^{3}}.  \label{R}
\end{equation}
Substituting the metric associated with the line element (\ref{ani}) into
the Klein Gordon equation (\ref{KG}) one obtains 
\begin{equation}
e^{2z}\frac{\partial ^{2}\Phi }{\partial x^{2}}+e^{2z}\frac{\partial
^{2}\Phi }{\partial y^{2}}-2\frac{\partial \Phi }{\partial z}+\frac{\partial
^{2}\Phi }{\partial z^{2}}-2\frac{\partial \Phi }{\partial \eta }\frac{%
d\alpha }{d\eta }\frac{1}{\alpha ^{3}}-e^{2z}A_{1}(y)^{2}\Phi -2ie^{2z}\frac{%
\partial \Phi }{\partial x}A_{1}(y)-M^{2}\alpha ^{2}\Phi =0,
\end{equation}
where we have chosen to work with a minimal coupling $\xi =0$. The Klein
Gordon equation (\ref{KG}) is completely separable in (\ref{ani}), therefore
we look for its solution in the form. 
\begin{equation}
\Phi =H(\eta )Z(z)Y(y)e^{ik_{x}x}.  \label{ans}
\end{equation}
Substituting (\ref{ans}) into Eq. (\ref{KG}) we reduce the problem of
solving the Klein-Gordon equation to that of finding solutions of the
following set of ordinary differential equations 
\begin{equation}
\frac{d^{2}Y}{dy^{2}}-\left( (k_{x}-A_{1}(y))^{2}-k^{2}\right) Y=0,  \label{Y}
\end{equation}
\begin{equation}
\frac{d^{2}Z}{dz^{2}}-2\frac{dZ}{dz}-(\lambda ^{2}+k^{2}e^{2z})Z=0,  \label{Z}
\end{equation}
\begin{equation}
\frac{d^{2}H}{d\eta ^{2}}+2\frac{dH}{d\eta }\frac{d\alpha }{d\eta }+(\alpha
^{2}(\eta )M^{2}-\lambda ^{2})H=0,  \label{H}
\end{equation}
with $\lambda ^{2}$ and $k^{2}$ as separation constants. \ For $A(y)=A_{1}y$
the solution of Eq. (\ref{Y}) can be expressed in terms of Whittaker
functions \cite{Abramowitz} as follows 
\begin{equation}
Y=C_{1}v^{-1/2}M_{\frac{k^{2}}{4A_{1}},\frac{1}{4}}(v^{2})+C_{2}v^{-1/2}W_{%
\frac{k^{2}}{4A_{1}},\frac{1}{4}}(v^{2}),  \label{Y1}
\end{equation}
where 
\begin{equation}
v=\frac{A_{1}y-k_{x}}{\sqrt{A_{1}}},  \label{nu}
\end{equation}
and $C_{1}$ and $C_{2}$ are arbitrary constants. In the absence of
electromagnetic field the solution of Eq. (\ref{Y}) reduces to 
\begin{equation}
Y=C_{1}e^{\pm i\sqrt{k^{2}-k_{x}^{2}}y}=C_{1}e^{\pm ik_{y}y}.  \label{Y2}
\end{equation}
The solution of Eq. (\ref{Z}) is \cite{Lebedev} 
\begin{equation}
Z=C_{3}e^{z}H_{\sqrt{1+\lambda ^{2}}}^{(1)}(ike^{z})+C_{4}e^{z}H_{\sqrt{%
1+\lambda ^{2}}}^{(2)}(ike^{z}),  \label{Zsol}
\end{equation}
where $H_{\nu }^{(1)}(z)$ and $H_{\nu }^{(2)}(z)$ are the Hankel functions
and $C_{3}$ and \ $C_{4}$ are arbitrary constants. We can also express the
solution of (\ref{Z}) in terms of Bessel functions $J_{\nu }(z)$ as 
\begin{equation}
Z=D_{3}e^{z}J_{\sqrt{1+\lambda ^{2}}}(ike^{z})+D_{4}e^{z}J_{-\sqrt{1+\lambda
^{2}}}(ike^{z})  \label{Zsol2}
\end{equation}
where $D_{3}$ and \ $D_{4}$ are arbitrary constants.

After introducing the function $h(\eta )$: 
\begin{equation}
H(\eta )=\frac{h(\eta )}{\alpha (\eta )},
\end{equation}
Eq. (\ref{H}) reduces to 
\begin{equation}
\frac{d^{2}h}{d\eta ^{2}}+(\alpha ^{2}(\eta )M^{2}-\lambda ^{2}-\frac{%
d^{2}\alpha (\eta )}{d\eta ^{2}})h=0.  \label{h}
\end{equation}
In order to analyze the asymptotic behavior of the solutions of the
Klein-Gordon equation (\ref{KG}) we make use of the asymptotic behavior of
the Hankel functions \cite{Abramowitz}

\begin{equation}
H_{\nu }^{(1)}(z)\rightarrow \left( \frac{2}{\pi z}\right) ^{1/2}e^{i(z-\pi
\nu /2-\pi /4)},H_{\nu }^{(2)}(z)\rightarrow \left( \frac{2}{\pi z}\right)
^{1/2}e^{-i(z-\pi \nu /2-\pi /4)},  \label{Hasy}
\end{equation}
as $z\rightarrow \infty $, and the behavior of $J_{\nu }(z)$ as $z\rightarrow
0$ \cite{Lebedev} 
\begin{equation}
J_{\nu }(z)\rightarrow \frac{\left( \frac{z}{2}\right) ^{\nu }}{\Gamma
(1+\nu )}  \label{J}
\end{equation}
The asymptotic behavior of the Whittaker function $W_{k,\mu }(z)$ for large
values of $z$ is \cite{Lebedev} 
\begin{equation}
W_{k,\mu }(z)\rightarrow e^{-z/2}z^{k},  \label{Wkm}
\end{equation}
and the function $M_{k,\mu }(z)$ has the following asymptotic behavior as $%
z\rightarrow 0$ 
\begin{equation}
M_{k,\mu }(z)\rightarrow e^{-z/2}z^{\mu +\frac{1}{2}}.  \label{Mkm}
\end{equation}
An approximate solution of Eq. (\ref{h}) can be obtained provided that the
expansion parameter $\alpha (\eta )$ satisfies the conditions of validity of
the adiabatic approximation. In this case one has that $h(\eta )$ has the
form \cite{Fulling,Parker69,Parker74} 
\begin{equation}
h(\eta )=\frac{1}{\sqrt{2W(\eta )}}\exp (\pm i\int^{\eta }W(\eta ^{^{\prime
}})d\eta ^{^{\prime }},
\end{equation}
with 
\begin{equation}
W(\eta )^{2}=\omega (\eta )^{2}\left[ 1+\delta _{2}(\eta )\omega ^{-2}+...%
\right],
\end{equation}
where the function $\omega (\eta )$ has the form 
\begin{equation}
\omega (\eta )^{2}=\alpha ^{2}(\eta )M^{2}-\lambda ^{2}-\frac{d^{2}\alpha
(\eta )}{d\eta ^{2}},
\end{equation}
$\delta _{n}(\eta )$ is a function of $\omega (\eta )$ and its derivatives
at $\eta $ up through $\omega ^{(n)}(\eta )$ and $\delta _{n}(\eta )$ \ is
bounded as $\omega \rightarrow \infty .$ The solution of the Klein-Gordon
equation (\ref{KG}) can be written as 
\begin{equation}
\Phi =\frac{\exp (\pm i\int \sqrt{\alpha ^{2}(\eta )M^{2}-\lambda ^{2}-\frac{%
d^{2}\alpha (\eta )}{d\eta ^{2}}}d\eta }{\sqrt{2}a(\eta )^{3/2}}%
Z(z)Y(y)e^{ik_{x}x}.  \label{solu}
\end{equation}
Let us analyze the asymptotic behavior of (\ref{solu}) as $y\rightarrow
\infty $ and $z\rightarrow -\infty $. Using (\ref{Sz}) and (\ref{J}) \ we
obtain that, when the electromagnetic interaction is switched off, the mode
solutions of Eq. (\ref{KG}) take the asymptotic form 
\begin{equation}
\Phi \rightarrow \frac{\exp (\pm i\int \sqrt{\alpha ^{2}(\eta )M^{2}-\lambda
^{2}-\frac{d^{2}\alpha (\eta )}{d\eta ^{2}}}d\eta }{\sqrt{2}a(\eta )^{3/2}}%
e^{\mp ik_{y}y}e^{ik_{x}x}e^{(\mp \sqrt{1+\lambda ^{2}}+1)z}.  \label{solu2}
\end{equation}
Analogously, we have that in the presence of the electromagnetic potential
the mode solutions of Eq. (\ref{KG}) take the
following asymptotic form 
\begin{equation}
\Phi \rightarrow \frac{\exp (\pm i\int \sqrt{\alpha ^{2}(\eta )M^{2}-\lambda
^{2}-\frac{d^{2}\alpha (\eta )}{d\eta ^{2}}}d\eta }{\sqrt{2}a(\eta )^{3/2}}%
e^{(\pm \sqrt{1+\lambda ^{2}}+1)z}\nu ^{k^{2}/2A_{1}}e^{-\nu
^{2}/2}e^{ik_{x}x}.  \label{nega}
\end{equation}
For large positive values of $z$ we have that the asymptotic behavior of $%
\Phi $ is 
\begin{equation}
\Phi \rightarrow \frac{\exp (\pm i\int \sqrt{\alpha ^{2}(\eta )M^{2}-\lambda
^{2}-\frac{d^{2}\alpha (\eta )}{d\eta ^{2}}}d\eta }{\sqrt{2}a(\eta )^{3/2}}%
e^{\mp e^{-ke^{z}}}e^{z/2}\nu ^{k^{2}/2A_{1}}e^{-\nu ^{2}/2}e^{ik_{x}x}.
\label{posi}
\end{equation}
From Eq. (\ref{nega}) we can identify the 
quasiclassical modes as $y\rightarrow \infty $ and $z\rightarrow -\infty $
as 
\begin{equation}
\Phi _{class(z\rightarrow -\infty )}=\frac{h(\eta )}{\alpha (\eta )}%
e^{z}J_{\pm \sqrt{1+\lambda ^{2}}}(ike^{z})v^{-1/2}W_{\frac{k^{2}}{4A_{1}},%
\frac{1}{4}}(v^{2})e^{ik_{x}x}  \label{mod}.
\end{equation}
Analogously, from Eq. (\ref{posi}) we have that the quasiclassical modes as $%
y\rightarrow \infty $ and $z\rightarrow \infty $ are 
\begin{equation}
\Phi _{class(z\rightarrow \infty )}=\frac{h(\eta )}{\alpha (\eta )}%
e^{z}H_{\pm \sqrt{1+\lambda ^{2}}}^{(1,2)}(ike^{z})v^{-1/2}W_{\frac{k^{2}}{%
4A_{1}},\frac{1}{4}}(v^{2})e^{ik_{x}x}.  \label{mod2}
\end{equation}

\section{Solution of the Dirac equation}

The Dirac equation is a system of coupled partial differential equations
which is separable in a very restricted set of metrics. Among the spacetimes
where the separability of the Klein-Gordon and Dirac equations has been
studied one can mention the St\"{a}ckel spaces \cite{Stackel}, which are
those metrics where the Hamilton-Jacobi equation is separable. Nevertheless
recently it has been shown that this condition is neither necessary nor
sufficient in order to guarantee a complete separability of variables in the
Dirac equation (see Ref. \cite{Varaksin} and references therein). A systematic
classification of the gravitational backgrounds where the Dirac equation is
separable with the help of the algebraic method is presented in ref. \cite
{Shishkin}. The line element (\ref{ani}) belongs to this family and
consequently one can apply the algebraic method of separation.

The covariant generalization of the Dirac equation in curved space-time is 
\cite{Birrel,Brill}

\begin{equation}
\tilde{\gamma}^{\alpha }\left( \partial _{\alpha }-\Gamma _{\alpha
}-ieA_{\alpha }\right) \tilde{\Psi}+M\tilde{\Psi}=0,  \label{D1}
\end{equation}
where the curved Dirac matrices $\tilde{\gamma}^{\alpha }$ satisfy the
commutation relation 
\begin{equation}
\left\{ \tilde{\gamma}^{\alpha },\tilde{\gamma}^{\beta }\right\} =2g^{\alpha
\beta },  \label{D2}
\end{equation}
and $\Gamma _{\alpha }$ are the spin connections \cite{Brill} 
\begin{equation}
\Gamma _{\alpha }=\frac{1}{4}g_{\mu \lambda }\left[ \left( \frac{\partial
b_{\nu }^{\beta }}{\partial x^{\alpha }}\right) a_{\beta }^{\lambda }-\Gamma
_{\nu \alpha }^{\lambda }\right] s^{\mu \nu },  \label{D3}
\end{equation}
where 
\begin{equation}
s^{\mu \nu }=\frac{1}{2}\left( \tilde{\gamma}^{\mu }\tilde{\gamma}^{\nu }-%
\tilde{\gamma}^{\nu }\tilde{\gamma}^{\mu }\right),  \label{D4}
\end{equation}
and the matrices\ $b_{\mu }^{\alpha }$, $a_{\beta }^{\mu }$ establish the
connection between the Dirac matrices $\tilde{\gamma}^{\mu }$ on a curved
space-time and the flat Dirac matrices $\gamma ^{\mu }$ as follows: 
\begin{equation}
\tilde{\gamma}_{\mu }=b_{\mu }^{\alpha }\gamma _{\alpha },\quad \tilde{\gamma%
}^{\mu }=a_{\beta }^{\mu }\gamma ^{\beta }.  \label{D5}
\end{equation}
Since the line element (\ref{ani}) is associated with a diagonal metric, we
can work in the diagonal tetrad gauge for $\tilde{\gamma}^{\mu }$: 
\begin{equation}
\tilde{\gamma}^{0}=\frac{\gamma ^{0}}{a(\eta )},\,\,\tilde{\gamma}^{1}=\frac{%
\gamma ^{1}}{a(\eta )e^{-z}},\,\,\tilde{\gamma}^{2}=\frac{\gamma ^{2}}{%
a(\eta )e^{-z}},\tilde{\gamma}^{3}=\frac{\gamma ^{0}}{a(\eta )},\,\,
\label{D6}
\end{equation}
substituting (\ref{D6}) into (\ref{D3}) we obtain that the spinor
connections are 
\begin{equation}
\Gamma _{1}=-\frac{1}{2}\frac{e^{-z}}{\alpha (\eta )}\left\{ -\alpha (\eta
)\gamma ^{1}\gamma ^{3}+\frac{d\alpha (\eta )}{d\eta }\gamma ^{1}\gamma
^{4}\right\},  \label{D7}
\end{equation}
\begin{equation}
\Gamma _{2}=-\frac{1}{2}\frac{e^{-z}}{\alpha (\eta )}\left\{ -\alpha (\eta
)\gamma ^{2}\gamma ^{3}+\frac{d\alpha (\eta )}{d\eta }\gamma ^{2}\gamma
^{4}\right\},  \label{D8}
\end{equation}
\begin{equation}
\Gamma _{3}=-\frac{1}{2}\frac{d\alpha (\eta )}{d\eta }\frac{1}{\alpha (\eta )%
}\gamma ^{3}\gamma ^{4},\quad \Gamma _{4}=0.  \label{D8a}
\end{equation}
Substituting (\ref{D6})-(\ref{D8a}) into (\ref{D1}) we find that the Dirac
equation takes the simple form 
\begin{equation}
\left\{ \gamma ^{0}\frac{\partial }{\partial \eta }+\gamma ^{1}e^{z}\left( 
\frac{\partial }{\partial x}-A_{1}(y)\right) +\gamma ^{2}e^{z}\frac{\partial 
}{\partial y}+\gamma ^{3}\frac{\partial }{\partial z}+M\alpha (\eta
)\right\} \Psi =0,  \label{equa}
\end{equation}
where we have introduced the spinor $\Psi $ 
\begin{equation}
\tilde{\Psi}=a(\eta )^{-3/2}e^{z}\Psi.  \label{D10}
\end{equation}
Regarding Eq. (\ref{equa}) we should mention that it does exhibits a 
nonfactorizable structure \cite{Villalba2,Fels}. 
In order to solve Eq. (\ref{equa}) we
apply the algebraic method of separation of variables \cite
{Shishkin,Villalba1,Villalba2,Hounkonnou,Hounkonnou2}. The method consists
in rewriting the Dirac equation (\ref{equa}) as a sum of two first order
differential operators $\hat{K}_{1},\hat{K}_{2}$ satisfying the relation 
\begin{equation}
\left[ \hat{K}_{1},\hat{K}_{2}\right] _{-}=0,\quad \left\{ \hat{K}_{1}+\hat{K%
}_{2}\right\} \Phi =0  \label{D11}
\end{equation}
with 
\begin{equation}
\gamma ^{3}\gamma ^{0}\Psi =\Phi,  \label{D12}
\end{equation}
and 
\begin{equation}
\hat{K}_{1}(x,y)\Phi =\left\{ \gamma ^{2}\frac{\partial }{\partial y}+\gamma
^{1}\left( \frac{\partial }{\partial x}-iA_{1}(y)\right) \right\} \gamma
^{3}\gamma ^{0}\Phi =ik\Phi,  \label{D13}
\end{equation}
\begin{equation}
\hat{K}_{2}(z,\eta )\Phi =e^{z}\left\{ \gamma ^{0}\frac{\partial }{\partial
\eta }+\gamma ^{3}\frac{\partial }{\partial z}+M\alpha (\eta )\right\}
\gamma ^{3}\gamma ^{0}\Phi =-ik\Phi.  \label{D14}
\end{equation}
It should be noticed that\ using the pairwise scheme of separation \ one has
been able to reduce the problem of solving the Dirac equation to finding
solutions of the decoupled system of Eqs. (\ref{D13}) and (\ref{D14}).
A further problem arises when we try to separate variables in Eq. (\ref{D14}%
). Here it is not possible to reduce the problem to a set of two commuting
first order differential operators. In order to
separate variables in Eq. (\ref{D14}) we re-write it in the following form: 
\cite{Villalba2,Villalba3}

\begin{equation}
\left( \hat{L}_{1}\gamma ^{3}\gamma ^{0}+\hat{L}_{2}\right) \Phi =0,
\label{D15}
\end{equation}
where $\hat{L}_{1}$ and $\hat{L}_{2}$ are two commuting differential
operators given by the expressions 
\begin{equation}
\hat{L}_{1}=\gamma ^{0}\frac{\partial }{\partial \eta }+M\alpha (\eta ),
\label{D17}
\end{equation}
\begin{equation}
\hat{L}_{2}=\gamma ^{0}\frac{\partial }{\partial z}+ike^{z},  \label{D18}
\end{equation}
In order to separate variables in Eq. (\ref{D15}) we introduce  the auxiliary
spinor ${\mathcal{Y}}$
\begin{equation}
\left( \hat{L}_{1}\gamma ^{3}\gamma ^{0}+\tilde{L}_{2}\right) {\mathcal{Y}}%
=\Phi,   \label{D16}
\end{equation}
where the differential operator $\tilde{L}_{2}$ is given by the expression 
\begin{equation}
\tilde{L}_{2}=\gamma ^{0}\frac{\partial }{\partial z}-ike^{z}.  \label{e1}
\end{equation}
Substituting (\ref{D16}) into (\ref{D15}) we obtain that ${%
\mathcal{Y}}$ satisfies the following equation 
\begin{equation}
\left\{ \hat{M}_{1}+\hat{M}_{2}\right\} {\mathcal{Y}}=0,
\end{equation}
with $\left[ \hat{M}_{1},\hat{M}_{2}\right] =0$, and 
\begin{equation}
\left( \hat{M}_{1}+\tilde{\lambda}\right) {\mathcal{Y}}=\left( -\frac{%
\partial ^{2}}{\partial z^{2}}-i\gamma ^{0}ke^{z}+k^{2}e^{2z}+\tilde{\lambda}%
\right) {\mathcal{Y}}=0,  \label{e2}
\end{equation}
\begin{equation}
\left( \hat{M}_{2}-\tilde{\lambda}\right) {\mathcal{Y}}=\left( \frac{%
\partial ^{2}}{\partial \eta ^{2}}+\gamma ^{0}M\frac{d\alpha (\eta )}{d\eta }%
+M^{2}\alpha ^{2}(\eta )-\tilde{\lambda}\right) {\mathcal{Y}}=0,  \label{e3}
\end{equation}
where\ $\tilde{\lambda}$ is a separation constant. \ Introducing the new
variable $u=2ke^{z}$, we have that Eq. (\ref{e2}) can be written as 
\begin{equation}
\left( \frac{\partial ^{2}}{\partial u^{2}}+\frac{i}{2u}\gamma ^{0}-\frac{1}{%
4}+(\frac{1}{4}-\lambda )\frac{1}{u^{2}}\right) {\mathcal{S}}=0,  \label{e4}
\end{equation}
where 
\begin{equation}
u^{-1/2}{\mathcal{S}}={\mathcal{Y}}.  \label{S}
\end{equation}
Choosing the following representation of the Dirac matrices \cite{Jauch} 
\begin{equation}
\gamma ^{0}=\left( 
\begin{array}{cc}
-i & 0 \\ 
0 & i
\end{array}
\right) ,\quad \gamma ^{j}=\left( 
\begin{array}{cc}
0 & \sigma ^{j} \\ 
\sigma ^{j} & 0
\end{array}
\right) ,\quad 1\leq j\leq 3  \label{repre}
\end{equation}
we readily obtain that the spinor $\Phi $ has the following structure 
\begin{equation}
\left[ \sigma _{1}\frac{\partial }{\partial y}-i\sigma _{2}(k_{x}-A_{1}(y))%
\right] \Phi _{1}=ik\Phi _{2},  \label{fi1}
\end{equation}
\begin{equation}
\left[ -\sigma _{1}\frac{\partial }{\partial y}+i\sigma _{2}(k_{x}-A_{1}(y))%
\right] \Phi _{2}=ik\Phi _{1},  \label{fi2}
\end{equation}
\begin{equation}
\Phi =\left( 
\begin{array}{c}
\Phi _{1} \\ 
\Phi _{2}
\end{array}
\right) =\left( 
\begin{array}{c}
\phi (y) \\ 
F\sigma ^{3}\phi (y)
\end{array}
\right) \exp (ik_{x}x),  \label{phi}
\end{equation}
where 
\begin{equation}
\phi (y)=\left( 
\begin{array}{c}
A(y) \\ 
B(y)
\end{array}
\right).   \label{phi2}
\end{equation}
Using the representation (\ref{repre}) we obtain that the solution of Eq. (%
\ref{e4}) can be written in terms of Whittaker functions 
\begin{equation}
S_{1,2}=D_{1}W_{-1/2,\sqrt{\lambda }}(u)+D_{2}M_{-1/2,\sqrt{\lambda }%
}(u),S_{3,4}=D_{3}W_{1/2,\sqrt{\lambda }}(u)+D_{4}M_{1/2,\sqrt{\lambda }}(u)
\label{Ssol}
\end{equation}
where $D_{1},$ $D_{2},$ $D_{3},$ $D_{4}$ do not depend on the variable $u$.
Looking at (\ref{e3}) and (\ref{e4}) we have that, for regular solutions at $u=0,
$ the spinor ${\mathcal{Y}}$ has the following structure: 
\begin{equation}
{\mathcal{Y}}=\left( 
\begin{array}{c}
a(y)c_{1}(\eta )u^{-1/2}M_{+\frac{1}{2},\sqrt{\lambda }}(u) \\ 
b(y)c_{1}(\eta )u^{-1/2}M_{+\frac{1}{2},\sqrt{\lambda }}(u) \\ 
c(y)c_{2}(\eta )u^{-1/2}M_{-\frac{1}{2},\sqrt{\lambda }}(u) \\ 
d(y)c_{2}(\eta )u^{-1/2}M_{-\frac{1}{2},\sqrt{\lambda }}(u)
\end{array}
\right) \exp (ik_{x}x),  \label{ysol}
\end{equation}
Substituting (\ref{ysol}) into (\ref{D16}) and noticing that \ Eq. (\ref{e3}%
) is equivalent to the following system of equations 
\begin{equation}
\left( \frac{\partial }{\partial \eta }-iM\alpha (\eta )\right) c_{1}(\eta )=%
\sqrt{\tilde{\lambda}}c_{2}(\eta ),  \label{c1}
\end{equation}
\begin{equation}
\left( \frac{\partial }{\partial \eta }+iM\alpha (\eta )\right) c_{2}(\eta )=%
\sqrt{\tilde{\lambda}}c_{1}(\eta ),  \label{c2}
\end{equation}
we obtain that the spinor $\Phi $ has the following structure 
\begin{equation}
\Phi =\left( 
\begin{array}{c}
A(v)c_{1}(\eta )e^{-z/2}M_{-\frac{1}{2},\sqrt{\tilde{\lambda}}}(2ke^{z}) \\ 
B(v)c_{1}(\eta )e^{-z/2}M_{-\frac{1}{2},\sqrt{\tilde{\lambda}}}(2ke^{z}) \\ 
iA(v)c_{2}(\eta )e^{-z/2}M_{\frac{1}{2},\sqrt{\tilde{\lambda}}}(2ke^{z}) \\ 
-iB(v)c_{2}(\eta )e^{-z/2}M_{\frac{1}{2},\sqrt{\tilde{\lambda}}}(2ke^{z})
\end{array}
\right) \exp (ik_{x}x),  \label{s1}
\end{equation}
where $A(v)$ and $B(v)$ satisfy the system coupled system of equations

\begin{equation}
\left( \frac{d}{dy}-(k_{x}-A_{1}(y))\right) B=ikA,  \label{A1}
\end{equation}
\begin{equation}
\left( \frac{d}{dy}+(k_{x}-A_{1}(y))\right) A=ikB,  \label{A2}
\end{equation}
where $v$ was defined in Eq. (\ref{nu}).

The corresponding solution of Eq. (\ref{D11}) in terms of the Whittaker
functions $W_{k,\mu }(z)$ has the form
\begin{equation}
\Phi =\left( 
\begin{array}{c}
\sqrt{\tilde{\lambda}}iA(v)c_{1}(\eta )e^{-z/2}W_{-\frac{1}{2},\sqrt{\tilde{%
\lambda}}}(2ke^{z}) \\ 
-i\sqrt{\tilde{\lambda}}B(v)c_{1}(\eta )e^{-z/2}W_{-\frac{1}{2},\sqrt{\tilde{%
\lambda}}}(2ke^{z}) \\ 
A(v)c_{2}(\eta )e^{-z/2}W_{\frac{1}{2},\sqrt{\tilde{\lambda}}}(2ke^{z}) \\ 
B(v)c_{2}(\eta )e^{-z/2}W_{\frac{1}{2},\sqrt{\tilde{\lambda}}}(2ke^{z})
\end{array}
\right) \exp (ik_{x}x).  \label{s2}
\end{equation}
Let us look for solutions of the system (\ref{A1}) and (\ref{A2}) when the
electromagnetic potential has the simple functional dependence $%
A_{1}(y)=A_{1}y.$ In this case one can obtain exact solutions for $A(v)$ and 
$B(v)$ in terms of hypergeometric functions. After making the change of
variable (\ref{nu}) and using the recurrence relations \cite{Abramowitz} 
\begin{equation}
(b-1)M(a,b-1,z)=(b-1)M(a,b,z)+z\frac{dM(a,b,z)}{dz},
\end{equation}
\begin{equation}
\frac{1}{a}\frac{dM(a,b,z)}{dz}+M(a,b,z)=M(a+1,b,z),
\end{equation}
\begin{equation}
\frac{dU(a,b,z)}{dz}-U(a,b,z)=-U(a,b+1,z),
\end{equation}
we find that the general solution of the system of equations (\ref{A1}) and (%
\ref{A2}) reads 
\begin{equation}
A=\frac{\sqrt{2A_{1}}}{ik}e^{-\frac{1}{2}v^{2}}(C_{1}M(-\frac{k^{2}}{4A_{1}}+%
\frac{1}{2},\frac{1}{2},v^{2})+C_{2}U(-\frac{k^{2}}{4A_{1}}+\frac{1}{2},%
\frac{1}{2},v^{2})),  \label{Ad}
\end{equation}
\begin{equation}
B=e^{-\frac{1}{2}v^{2}}v(C_{1}M(-\frac{k^{2}}{4A_{1}}+\frac{1}{2},\frac{3}{2}%
,v^{2})-C_{2}U(-\frac{k^{2}}{4A_{1}}+\frac{1}{2},\frac{3}{2},v^{2})).
\label{Bd}
\end{equation}
The exact solution of the system of equations (\ref{A1})-(\ref{A2}) in the
absence of electromagnetic interaction has the form 
\begin{equation}
A=C_{1}e^{i\sqrt{k^{2}-k_{x}^{2}}y}+C_{2}e^{-i\sqrt{k^{2}-k_{x}^{2}}y},
\end{equation}
\begin{equation}
B=\frac{\sqrt{k^{2}-k_{x}^{2}}-ik_{x}}{k}C_{1}e^{i\sqrt{k^{2}-k_{x}^{2}}y}-%
\frac{\sqrt{k^{2}-k_{x}^{2}}+ik_{x}}{k}C_{2}e^{-i\sqrt{k^{2}-k_{x}^{2}}y},
\end{equation}
where $C_{1}$ and $C_{2}$ are arbitrary constants. The solutions of the
Dirac equation (\ref{s1}) and (\ref{s2}) exhibit an asymptotic behavior
which can be identified with the quasiclassical solutions of the
Hamilton-Jacobi equation (\ref{3}). With the help of the asymptotic
expressions (\ref{Mkm}), we find that the Dirac spinor $\Phi $ as $%
z\rightarrow -\infty $, and $y\rightarrow \infty $. takes the form 
\begin{equation}
\Phi _{z\rightarrow -\infty }=\left( 
\begin{array}{c}
\frac{\sqrt{2A_{1}}}{ik}c_{1}(\eta ) \\ 
-vc_{1}(\eta ) \\ 
-\frac{\sqrt{2A_{1}}}{k}c_{2}(\eta ) \\ 
ivc_{2}(\eta )
\end{array}
\right) e^{-ke^{z}}e^{\sqrt{\tilde{\lambda}}z}e^{-\frac{v^{2}}{2}}v^{\frac{%
k^{2}}{2A_{1}}-1}\exp (ik_{x}x),  \label{resh}
\end{equation}
where the functions $c_{1}(\eta )$ and $c_{2}(\eta )$ \ satisfy the system
of equations (\ref{c1}) and (\ref{c2}). For asymptotically large values of $z$
we have that the spinor $\Phi $ takes the form.   
\begin{equation}
\Phi _{z\rightarrow \infty }=\left( 
\begin{array}{c}
\sqrt{\tilde{\lambda}}\frac{\sqrt{2A_{1}}}{k}c_{1}(\eta )e^{-z} \\ 
i\sqrt{\tilde{\lambda}}vc_{1}(\eta )e^{-z} \\ 
-i2\sqrt{2A_{1}}c_{2}(\eta ) \\ 
-2kvc_{2}(\eta )
\end{array}
\right) e^{-ke^{z}}e^{-\frac{v^{2}}{2}}v^{\frac{k^{2}}{2A_{1}}}\exp (ik_{x}x).
\label{resh2}
\end{equation}
Looking at the solution of the Hamilton-Jacobi equation we can identify (%
\ref{s1}) and (\ref{s2}) as the corresponding quasiclassical modes as  $%
z\rightarrow -\infty $ and $z\rightarrow \infty $, respectively.  An
approximate expression for the time dependence of the spinor $\Phi $ can be
obtained with the help of the WKB approximation. In this case we obtain 
\begin{equation}
c_{2}(\eta )\sim c_{10}\exp (i\omega (\eta )),  \label{c1a}
\end{equation}
\begin{equation}
c_{1}(\eta )\sim -i\frac{c_{10}}{\omega (\eta )+M\alpha (\eta )}\exp
(i\omega (\eta )),  \label{c1b}
\end{equation}
where $c_{10}$ is a normalization constant and $\omega (\eta )=\sqrt{iM\frac{%
d\alpha }{d\eta }+M^{2}\alpha ^{2}-\tilde{\lambda}}$. Looking at (\ref{c1a}%
)-(\ref{c1b}) and (\ref{resh}) we readily see that, for large values of $%
\eta $ we obtain $c_{1}(\eta )\rightarrow -i\frac{c_{10}}{2M\alpha (\eta )}%
\exp (i\omega (\eta )).$ \ Analytic solutions of the system of equations (%
\ref{c1a}) and (\ref{c1b}) can be obtained for some particular expansion
parameter $\alpha (\eta )$ \cite{Villalba2,Villalba3}

\section{Concluding remarks}

In this article, we have solved the Klein-Gordon and Dirac equations in an
open cosmological universe with partially horn topology. The solutions of
the relativistic wave equations are expressed in terms of special functions.
In Sec. IV we have shown that the algebraic method of separation \cite
{Villalba2,Hounkonnou,Hounkonnou2,Villalba3} permits one a complete
separation of variables of the Dirac equation in the line element associated
with a horn topology. The identification of the quasiclassical modes with
the help of the relativistic Hamilton-Jacobi equation shows that this method
is a very useful tool in the study of quantum effects in curved spaces.

As a final remark, we should mention that the introduction of nonstandard
topologies in order to describe the large scale structure of the space-time
also opens new possibilities to discuss quantum effects in globally
inhomogeneous and
anisotropic backgrounds in the presence of non-trivial electromagnetic
interactions. 

\section*{acknowledgments}

We want to express our gratitude to the referee for his critical remarks.
One of the authors (VMV) acknowledges a fellowship from the Alexander von
Humboldt Stiftung.

\end{document}